\documentclass[prd,twocolumn,secnumarabic]{revtex4}
\usepackage{amsmath,graphicx,supertabular}

\begin{document}

\newcommand{\weff}{w_\text{eff}}
\def\lsim{\,\hbox{\lower0.6ex\hbox{$\sim$}\llap{\raise0.6ex\hbox{$<$}}}\,}

   \title{Probing dark energy with future surveys}
\author{Roberto Trotta}
   \email{rxt@astro.ox.ac.uk} \affiliation{Astrophysics Department,
University of Oxford, Denys Wilkinson Building, Keble Road, Oxford
OX1 3RH, UK}

   \begin{abstract}
   I review the observational prospects to constrain the equation of
state parameter of dark energy and I discuss the potential of
future imaging and redshift surveys. Bayesian model selection is
used to address the question of the level of accuracy on the
equation of state parameter that is required before explanations
alternative to a cosmological constant become very implausible. I
discuss results in the prediction space of dark energy models. If
no significant departure from $w=-1$ is detected, a precision on
$w$ of order 1\% will translate into strong evidence against
fluid--like dark energy, while decisive evidence will require a
precision of order $10^{-3}$.
\end{abstract}

   \maketitle
%
%
\section{Introduction}

One of the most fundamental problems of contemporary physics is to
elucidate the nature of the ``Dark Sector'' of the Universe. A
wealth of cosmological observations seem presently to point to a
concordance cosmological model where ``normal'' (i.e.\ baryonic)
matter accounts for a mere 4\% of the matter--energy contents of
the cosmos. The remaining 96\% makes up the so--called ``Dark
Sector'', with about 19\% of cold dark matter (CDM) and 77\% of
``dark energy''. The details of this cosmic budget vary somewhat
depending on the data sets used and the assumptions one makes, but
the errors on the different components are below 10\% (for
details, see
e.g.~\cite{Efstathiou,Percival,Seljak,Sanchez,Spergel:2006hy}).

Guidance as to the nature of dark energy requires stronger
observational proof of its properties today and in the past. A
first important step is to discriminate between an evolving dark
energy (whose energy density changes with cosmic time) and a
cosmological constant of the form proposed by Einstein in the
1910s. A handle on this question is offered by the equation of
state parameter, $w$, that measures the ratio of pressure to
energy density of dark energy. Current data are consistent with
$w=-1$ out to a redshift of about $1$, with an uncertainty of
order $5-10\%$, by using all of the available data sets (see
e.g.~\cite{Seljak:2006bg}). However, one must be very careful when
assessing the combined constraining power of different data sets
whenever each one of them does not provide strong constraints when
taken alone. Combination of mutually inconsistent data can
potentially lead to unwarranted conclusions on the dark energy
parameters.

We first briefly review the observational prospects for
constraining the dark energy equation of state parameter,
referring the reader to \cite{Trotta:2006gx} for a more detailed
discussion. In section \ref{sec:bayes} we present and discuss some
results on the required accuracy on $w$ from the perspective of
Bayesian model selection and conclude in section
\ref{sec:conclusion}.

\section{What will we learn in the next decade?}

\label{sec:tech}

The observable impact of dark energy can be broadly divided in two
classes: modification of the redshift--distance relation and
effects on the growth of structures. Accordingly, we can divide
the different methods one can use to constrain dark energy
following the effect through which they are mainly sensitive to
dark energy: {\em probes of the redshift--distance relation}
(supernovae as standard candles, acoustic oscillations as standard
rulers) and {\em probes of the growth of structures} (galaxy
clustering, number counts, weak lensing, Integrated Sachs--Wolfe
effect). For a more detailed review of the advantages and
weaknesses of each technique, see  \cite{Trotta:2006gx}.

While none of those methods possess by itself all of the {\em
desiderata} that we would ideally want in trying to constrain dark
energy, combination of (at least) two techniques offers many
advantages. It allows for cross--calibration of observables and
facilites cross--checks of systematics, since the physical
underpinnings of each observable are different, and so is the
nature of the possible systematic errors.

In view of this, a very promising combination is given by weak
lensing and baryonic acoustic oscillations, which together offer
the advantages of potentially high accuracy (weak lensing) and
robustness to systematics (acoustic oscillations). They
independently probe the growth of structures (lensing) and the
angular diameter distance relation (acoustic oscillations, once
calibrated against the high--redshift ruler given by the cosmic
microwave background). Weak lensing studies will need
high--quality imaging surveys covering many thousands of square
degrees, while spectrographic redshift surveys encompassing
millions of galaxies will be necessary to exploit fully the
potential of acoustic oscillations. Let us now review the
observational perspectives in those fields over the next decade.

\subsection{Imaging surveys}

Proposals for the next generation of imaging surveys driven by
dark energy science typically feature a survey area covering 5,000
to 10,000 square degrees, a large field of view (2 square degrees
or more) and four to five optical photometric bands. Those are the
basic specifications for both the {\em Dark Energy Survey} (DES)
and {\em darkCAM}, which would have optical cameras mounted on 4m
class telescopes. DES is a US--led collaboration that will use a
520 megapixel CCD camera mounted on the Blanco telescope to image
300 million galaxies at a median redshift of $z \sim 0.7$ and to
carry out weak lensing, baryonic oscillations, cluster counts and
SNe observations over 5 years, starting in 2009. The European
UK--led {\em darkCAM} proposal to image some $10^9$ galaxies with
weak lensing image quality was originally envisaged to share time
on ESO's VISTA, but is now looking at a full--time site.

One of the most advanced projects is the {\em Pan-STARRS} survey
(Panoramic Survey Telescope and Rapid Response System), a US Air
Force funded project in Hawaii, primarily devoted to the
identification of Earth-approaching objects, but with 30\% of its
time dedicated to supernovae, baryon oscillations and weak lensing
surveys. The first of the planned four 1.8m telescopes is
currently undergoing commissioning, and the full system could be
online by about 2009, representing a major increase in power with
respect to present--day surveys.

In purely statistical terms, the most precise constraints on the
dark energy equation of state are likely to come from weak
lensing. The details depend very much on which assumptions are
made about the cosmology and on which other data sets are
included. By exploiting all of the correlations that can be
constructed from a weak lensing survey, weak lensing alone could
achieve better than 5\% accuracy on the effective equation of
state, while in combination with CMB anisotropies measurements of
Planck quality (an ESA satellite missione due for launch at the
beginning of 2008) an accuracy of 1--2\% might be within reach.
This is of course only achievable if all of the systematic errors
will be kept closely under control. This means an exquisite image
quality, good seeing conditions (below 0.9 arcsec), excellent
photometric redshift reconstruction and control of intrinsic and
gravitational--intrinsic correlations. Arguably, the major
hindrance in pushing weak lensing constraints below the 5\% mark
will indeed come from systematic error control.

The clusters and SNe method will be considerably less stringent,
roughly a factor of 3 to 4 less precise than weak lensing, unless
combined with strong CMB priors (i.e., Planck data), in which they
case they will perform at about the 5\% level. The performance of
the cluster count technique relies however on self--calibration
using clustering and weak lensing data, a difficult procedure
compounded by the challenge of controlling systematic errors at
this level of precision. The possibility of SNe evolution and
missing pieces in our understanding of how a supernova explosion
comes about are also likely to be limiting factors when trying to
increase the accuracy on the equation of state below the 10--5\%
limit with this technique. Finally, measurements of acoustic
oscillations from imaging surveys are not competitive with the
other methods in terms of precision, reaching down to only about
20\% accuracy because of the lack of resolving power in the radial
direction (\cite{Blake:2004tr}, but see also \cite{Angulo}).

\subsection{Spectrographic surveys}

There are a number of redshift surveys at various stages of
planning, development or commissioning, that will have among their
main science drivers measurements of the acoustic ruler at
different redshifts.

Perhaps the most ambitious is the {\em Wide-Field Multi-Object
Spectrograph (WFMOS)} (see \cite{Bassett:2005kn}), a proposal for
a 1.5 deg$^2$ multi-object spectrograph which will be able to
observe 4,000 to 5,000 objects simultaneously. The instrument is
to be developed collaboratively by the Gemini and Subaru
Observatories and will be deployed on the 8m Subaru telescope on
Mauna Kea, Hawaii. Two baseline surveys are being proposed: a
shallower and wider one, covering 2,000 square degrees at $z~\sim
1$ which will target emission line blue galaxies; and a deeper
one, over 300 square degrees at $z \sim 3$ targeting Lyman-Break
Galaxies. The two WFMOS proposed baseline surveys will determine
the angular diameter distance and the Hubble expansion rate at $z
\sim 1$ and $z \sim 3$ with 1--2\% accuracy. The corresponding
constraints on the dark energy equation of state rely on the
calibration of the acoustic scale. If combined with Planck
forecasts and SDSS data, WFMOS observations should achieve an
accuracy in the range of 5--10\% in the effective equation of
state.

On a shorter timescale, there are proposals to use the {\em
AAOmega} wide--field spectrograph -- an upgrade to the 2dF
spectrograph for the Ango--Australian Telescope, which has now
been successfully commissioned -- to carry out large surveys
(between 500 and 1,000 deg$^2$) in the redshift range $0.3 < z <
1$ to achieve 2\% accuracy in the angular diameter distance and
the expansion rate. A rather more revolutionary concept is being
investigated for the {\em VIRUS} spectrograph, a proposal for the
9m Hobby-Eberly Telescope in Texas based on industrial replication
of low--cost components.

In summary, the statistical accuracy from acoustic oscillations
redshift surveys is less than what could be achieved with weak
lensing. However, the acoustic oscillation method seems to be much
more robust with respect to systematic errors, and it can probe a
deeper redshift range than any other method.

\subsection{On the pathway to the SKA}

On a slightly longer timescale, proposals for next--to--next
generation of instruments aim at taking dark energy investigations
to an even more ambitious level. Among them, perhaps the most
prominent are the {\em LSST} in the optical and the {\em SKA} in
the radio.

The {\em LSST} (Large Synoptic Survey Telescope) is a project for
a wide--field, 8.4m telescope and a 3 Gpixels camera. The survey
will cover the whole of the Southern hemisphere (or 20,000
deg$^2$) multiple times per month with 6 colours photometry. It
will survey the largest volume ever proved and it will use a
variety of techniques (weak lensing, acoustic oscillations,
cluster abundance and a staggering 250,000 SNe per year) to
constrain dark energy at the percent level. The current schedule
expects construction to begin in 2009 and first light in 2013.
Science will start in 2014.

The second half of the next decade will also see a great leap
forward in radio astronomy, as the SKA (Square Kilometer Array)
begins operations, first as a pathfinder (around 2015) and then as
a full system with a total collecting area of a million square
meters (around 2020). Thanks to its huge field of view, the SKA
will be able to measure redshifts of a billion of galaxy over half
of the sky in only a few months of operations, by detecting radio
emissions from hydrogen gas (see eg \cite{Blake:2004pb}). The
project is now beginning the design study phase, thanks to a
recent funding decision by the European partners, including PPARC.

\section{How far should we go in establishing $w = -1$? \label{sec:bayes}}

If one could determine with high accuracy that $w=-1$ and constant
in time, this would strongly support the case for a cosmological
constant. This would imply that dark energy is a manifestation of
a new constant of Nature, whose magnitude would however suffer
from a strong fine tuning problem. Detecting an evolution with
redshift of $w(z)$ would support a dynamical form of dark energy,
perhaps in the form of a scalar field that could be linked to the
inflationary phase of the early Universe. Either one of these
results is likely to have a major impact on our knowledge of
fundamental physics.

Since current data are compatible with $w=-1$ at all redshifts $<
1$, it is interesting to ask what level of accuracy is required
before our degree of belief in the cosmological constant is
overwhelmingly larger than for an evolving dark energy. This of
course assumes that future data will not detect any significant
departure from $w = -1$. Bayesian model comparison is a
quantitative tool to address this question that takes into account
the predictivity of the more complicated model (in this case, a
time varying dark energy) and the information content of the data,
see \cite{Trotta:2005ar} and \cite{Kunz:2006mc} for a discussion
and general introduction. In this case, the relevant quantity is
the Bayes factor $B$ between a cosmological constant model ($w =
-1$) and a varying dark energy model with an effective equation of
state (averaged over redshift with the appropriate weighting
factor for the observable, see \cite{Simpson:2006bd}) $\weff \neq
-1$. The Bayes factor gives the amount by which our relative
believe in the two models is modified by the data, with $\ln B >
(<0)$ indicating a preference for the cosmological constant
(evolving dark energy) model. If we assume that the data are
compatible with $\weff=-1$ with an uncertainty $\sigma$, then the
Bayes factor in favour of a cosmological constant is given by
 \begin{equation} \label{eq:B}
 B = \sqrt{\frac{2}{\pi}}\frac{\Delta_{+} + \Delta_{-}}{\sigma}
 \left[\text{erfc}\left(-\frac{\Delta_+}{\sqrt{2}\sigma}\right)
- \text{erfc}\left(\frac{\Delta_-}{\sqrt{2}\sigma}\right)
  \right]^{-1},
 \end{equation}
where for the evolving dark energy model we have adopted a flat
prior in the region $-1 - \Delta_{-} \leq \weff \leq -1+\Delta_+$
and we have made use of the Savage--Dickey density ratio formula
(see \cite{Trotta:2005ar}). The prior, of total width $\Delta =
\Delta_+ + \Delta_-$, is best interpreted as a factor describing
the predictivity of the dark energy model under consideration. For
instance, in a model where dark energy is a fluid with a negative
pressure but satisfying the strong energy condition we have that
$\Delta_+ = 2/3, \Delta_- = 0$. On the other hand, phantom models
will be described by $\Delta_+ = 0, \Delta_- > 0$, with the latter
being possibly rather large (see e.g. \cite{Kujat:2006vj} for an
example). A model with a large $\Delta$ will be more generic and
less predictive, and therefore is disfavoured by the Occam's razor
of Bayesian model selection. We notice that the prior and its
flatness over the range $\Delta$ do not necessarily need to be
interpreted as reflecting a probability distribution of models in
terms of frequency of outcomes (although the latter could easily
be implemented if available) but rather our state of knowledge
about the range of possibilities that can be realized by the model
{\em a priori}. According to the Jeffreys' scale for the strength
of evidence, we have a strong (decisive) preference for the
cosmological constant model for $3.0 < \ln B < 5.0$ ($\ln B>5.0$),
corresponding to posterior odds of $20:1$ to $150:1$ (above
$150:1$). 
\begin{figure}[tb]
\centering
\includegraphics[width=\linewidth]{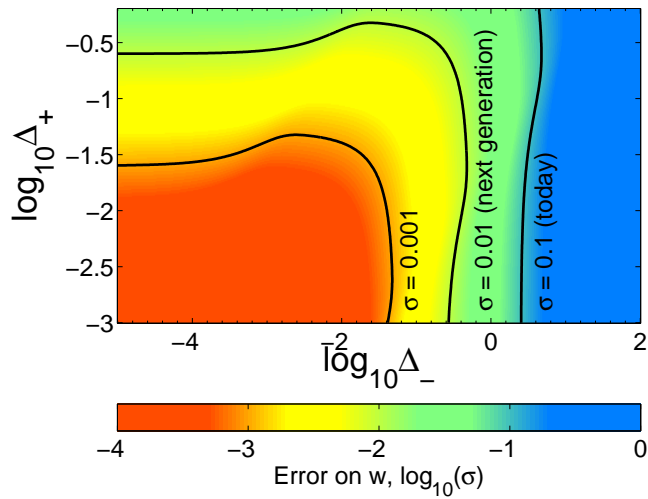}
\caption{Required accuracy on $\weff = -1$ to obtain strong
evidence against a model where $-1 - \Delta_{-} \leq \weff \leq
-1+\Delta_+$ as compared to a cosmological constant model, $w=-1$.
For a given $\sigma$, models to the right and above the contour
are disfavoured with strong (i.e. $>20:1$) odds. The benchmark
models discussed in the text are located in the upper left corner
(fluid--like dark energy, for $\log_{10} \Delta_- \rightarrow
-\infty$), bottom right axis (phantom dark energy, for $\log_{10}
\Delta_+ \rightarrow -\infty$) and at the coordinates $(-2,-2)$
(percent--level departures from $w=-1$). } \label{fig:sigma}
\end{figure}
\begin{table}
\begin{tabular}{l l|l }
 Model & $(\Delta_+ , \Delta_- )$ & $\ln B$ today ($\sigma = 0.1$)
 \\\hline
 Phantom & $( 0, 10)$      & $4.4$ (strongly disfavoured)\\
 Fluid--like & $(2/3,  0)$ & $1.7$ (slightly disfavoured) \\
 Small departures & $( 0.01, 0.01)$ & $0.0$ (inconclusive) \\
\end{tabular}
\caption{Strength of evidence disfavouring the three benchmark
models against a cosmological constant model, using an indicative
accuracy on $w=-1$ from present data, $\sigma \sim 0.1$.
\label{table:evidence}}
\end{table}

We plot in Fig.~\ref{fig:sigma} contours of constant observational
accuracy $\sigma$ in the model predictivity space
$(\Delta_-,\Delta_+)$ for $\ln B = 3.0$ from Eq.~\eqref{eq:B},
corresponding to strong evidence in favour of a cosmological
constant. The figure can be interpreted as giving the space of
extended models that can be significantly disfavoured with respect
to $w=-1$ at a given accuracy. Present--day precision, roughly of
order $\sigma \sim 10^{-1}$, gives odds stronger than $20:1$
against phantom models with $\Delta_- \lsim 1$ as compared to a
cosmological constant model. As we have seen, the next generation
of dark energy surveys will reach $\sigma \sim 0.01$ and this will
allow to strongly disfavour (or otherwise) fluid--like models,
corresponding to the top left corner of the figure. Another order
of magnitude increase in precision is required to test with strong
significance models which predict percent--level departures from
$w=-1$. The results for the 3 benchmark models mentioned above
(fluid--like, phantom or small departures from $w=-1$) are
summarized in Table~\ref{table:evidence}, where we list the
outcome of present--day model comparison against $w=-1$. Only
phantom models with large $\Delta_-$ are presently significantly
disfavoured.

In Table~\ref{table:sigma} we show the required accuracy in terms
of $\sigma$ in order to achieve strong evidence against each of
the models. We conclude that in the lack of significant departures
from $\weff=-1$  future surveys will be able to accumulate
decisive evidence against phantom models with large ($\Delta_- >
10$) negative effective equation of state. Disfavouring a
fluid--like model where $-1 \leq \weff \leq -1/3$ will instead
require better than percent accuracy on $\weff$ ($\sigma =
3\cdot10^{-3}$). If models can be constructed that naturally
predict only percent--level departures from $\weff$ (that we
termed ``small departure models''), then Bayesian model selection
will not be able to strongly disfavour them unless the error on
$\weff$ could be decreased well below $10^{-4}$.
\begin{table}
\begin{tabular}{l l|l c }
 Model & $(\Delta_+ , \Delta_- )$ & \multicolumn{2}{c}{Required $\sigma$ for evidence level} \\
       &                          & strong  & decisive  \\\hline
 Phantom & $( 0, 10)$      & $0.4$           & $5\cdot10^{-2}$\\
 Fluid--like & $(2/3,  0)$ & $3\cdot10^{-2}$ & $3\cdot10^{-3}$\\
 Small departures & $( 0.01, 0.01)$ & $4\cdot10^{-4}$ & $5\cdot10^{-5}$\\
\end{tabular}
\caption{Required accuracy for future surveys in order to
disfavour the three benchmark models against $w=-1$ with strong
($\ln B = 3$) or decisive ($\ln B = 5$) strength of evidence.
\label{table:sigma}}
\end{table}

We expect that a similar analysis could be easily carried out to
compare the cosmological constant model against departures from
Einstein gravity, thus giving some useful insight into the
potential of future surveys in terms of Bayesian model selection
(see also \cite{Mukherjee:2005tr} for a similar approach).

\section{Conclusions}
\label{sec:conclusion}

We have argued that the most promising methods for dark energy
investigation are weak lensing and acoustic oscillations, because
of their statistical accuracy (weak lensing) and robustness to
systematic errors (acoustic oscillations). Weak lensing has the
potential of achieving 1\% accuracy on $\weff$ but this precision
requires an exquisite control of various systematic errors.
Observations of baryonic oscillations with a spectroscopic survey
have less statistical power than weak lensing (roughly a factor of
5), but are less prone to systematic errors due to the
characteristics of the acoustic signature. The above goals could
be reached within the next decade thanks to a vigorous
observational campaign, involving collaborations such as DES,
darkCAM, Pan--STARRS and WFMOS.

We have shown that Bayesian model selection can offer a guidance
as to the level of precision required on $w=-1$ before
explanations alternative to a cosmological constant appear
extremely unlikely in terms of posterior odds of models. We have
found that phantom models where one can have $\weff \ll -1$ are
strongly disfavoured by present--day data. Gathering decisive
evidence against a fluid--like model for dark energy will however
require a precision of order $10^{-3}$ on $\weff$.

In conclusion, the observational study of dark energy is a crucial
area of cosmological research. Thanks to a host of ambitious
proposals and a strong support by several funding bodies, key
advances are likely to be made within the next decade both from
the observational and the theoretical points of view.

\begin{acknowledgements}
RT is supported by the Royal Astronomical Society through the Sir
Norman Lockyer Fellowship and thanks PPARC for support of this
work. I am grateful to Sarah Bridle, Ofer Lahav and Jochen Weller
for stimulating discussions.
\end{acknowledgements}

 \end{document}